\documentstyle[aps,prl,twocolumn,epsfig]{revtex}
\begin{document}
\draft
\title{Quantum Dynamics of Ultracold Na + Na$_2$ collisions}
\author{Pavel Sold\'{a}n, Marko T. Cvita\v s and Jeremy M. Hutson}
\address{Department of Chemistry, University of Durham, South Road,
Durham, DH1~3LE, England}
\author{Pascal Honvault and Jean-Michel Launay}
\address{UMR 6627 du CNRS, Laboratoire de Physique des Atomes, Lasers,
Mol\'ecules et Surfaces, Universit\'e de Rennes, France}
\date{\today}
\pacs{34.50.-s,03.75.Fi}
\maketitle
\begin{abstract}

Ultracold collisions between spin-polarized Na atoms and vibrationally
excited Na$_2$ molecules are investigated theoretically, using both an
inelastic formalism (neglecting atom exchange channels) and a reactive
formalism (including atom exchange). Calculations are carried out on both
pairwise additive and non-additive potential energy surfaces for the
quartet electronic state. In both inelastic and reactive calculations, the
Wigner threshold laws are followed for energies below $10^{-6}$~K. It is
found that vibrational relaxation processes dominate elastic processes for
temperatures below $10^{-3}-10^{-4}$~K. For temperatures below $10^{-5}$~K,
the rate coefficients for vibrational relaxation ($v=1\rightarrow 0$) from
the full calculation are $4.8\times10^{-11}$ and $5.2\times10^{-10}$ cm$^3$
s$^{-1}$ for the additive and non-additive potentials respectively.

\end{abstract}
\pacs{}

\font\smallfont=cmr7

Methods for creating diatomic molecules in atomic Bose-Einstein condensates
(BECs) are now starting to be realised experimentally. They include
photoassociation spectroscopy \cite{Stw99,Wyn00,Ger00,McK02} and magnetic
tuning through a Feshbach resonance \cite{Mie00,Don02}. An important
long-range goal of such experiments is the production of a stable molecular
BEC. However, in most cases the molecules are produced initially in
vibrationally excited states, and their lifetime is limited by collisions
with other atoms or molecules. Since the magnetic trap depth is typically 1
mK, any vibrationally or rotationally inelastic collision will release
enough kinetic energy for both collision partners to be ejected from the
trap.

Very little is known about vibrational relaxation rates for ultracold
alkali dimers. Wynar {\em et al.} \cite{Wyn00} produced $^{87}$Rb$_2$
molecules in very high vibrational levels of the ground electronic state in
an atomic BEC by stimulated Raman scattering. They analyzed their line
shapes to obtain an upper bound on the inelastic rate coefficient $k^{\rm
inel} < 8 \times 10^{-11}$ cm$^3$/s. In subsequent experiments \cite{Hei02}
on ultracold (but not condensed) Rb$_2$ molecules in a different set of
vibrational levels, they measured $k^{\rm inel} = 3 \times 10^{-11}$
cm$^3$/s. These rates are too high for the production of long-lived
molecules, but there is a hope that lower-lying states will relax more
slowly \cite{Bal97thresh,Bal98HeH2} and that methods for stabilizing the
molecules can be found.

Several calculations have been carried out on the vibrational relaxation of
molecules such as H$_2$ \cite{Bal97thresh,Bal98HeH2} and CO
\cite{Bal00HeCO,Zhu01HeCO} at ultralow energies. Spin-changing collisions
of O$_2$ have also been investigated. \cite{Boh00HeO2,Avd01O2O2} However,
collisions of alkali dimers present new theoretical challenges that are not
present for collisions of stabler molecules. In particular, the potential
energy surfaces are such that barrierless atom-exchange reactions can
occur; even if the products are indistinguishable from the reactants, the
reactive channels must be taken into account in a full treatment of the
collision dynamics. Barrierless reactions are significantly different from
reactions such as F + H$_2$, \cite{Bal01FH2} which have substantial
barriers, and have not yet been investigated at ultralow energies. This
Letter presents an initial investigation of alkali + alkali dimer
collisions, for the case of Na + Na$_2$ collisions occurring on the lowest
quartet surface for Na$_3$. This corresponds physically to collisions of
atoms in their ``stretched'' spin states, with $F=F_{\rm max}=I+S$ and
$|M_F|=F$.

In this work we use the potential energy surfaces of Higgins {\em et al.}
\cite{Hig00} for the $1^4A'_2$ state of Na$_3$. Their full (non-additive)
potential was obtained from a grid of {\it ab initio} calculations with a
large basis set, interpolated using the reproducing-kernel Hilbert space
scheme of Ho and Rabitz \cite{hojcp104}. The potential has a global minimum
at $-1222.1$~K, with the atoms in an equilateral triangle configuration
4.41 \AA\ apart. The symmetric linear geometry is a saddle point at
$-554.4$~K (667.7~K above the minimum), with $r=5.10$ \AA. Comparison with
the corresponding triplet Na$_2$ pair potential \cite{gutjcp110}, which has
$r_e=5.192$ \AA\ and $D_e=255.7$~K, shows that large non-additive effects
are present. In the present work, we test the sensitivity of our results
to the potential surface by performing calculations using both the full
non-additive potential and a potential obtained by neglecting the
non-additive part.

We initially carried out inelastic scattering calculations of vibrational
relaxation from the $v=1$ state of triplet Na$_2$, using the MOLSCAT
program \cite{molscat}. Such calculations are carried out in Jacobi
coordinates $(R,r,\theta)$ for a single arrangement of the atoms, and
neglect the reactive channels. The wavefunction in the interaction region
is expanded in a basis set of Na$_2$ vibrational functions that are
eigenfunctions of the Hamiltonian for free Na$_2$, supplemented by a
wall at large $R$ to provide a representation of the Na$_2$ continuum.

For full close-coupling or coupled states calculations, it proved
impossible to converge the basis set of Na$_2$ vibrational functions. This
may readily be understood in terms of the potential energy surface.
Consider Na approaching Na$_2$ for a T-shaped geometry, $\theta=90^\circ$,
allowing the Na$_2$ bond length $r$ to relax to minimize the energy at each
intermolecular distance $R$. The system passes through the minimum-energy
equilateral geometry, and then the two atoms of the Na$_2$ move apart to
allow the Na atom to insert between them. However, even at $R=0$ (the
linear geometry), the energy is below that for separated Na + Na$_2$. At
this configuration, the optimum value of $r$ is 10.4 \AA, and the basis set
of Na$_2$ vibrational functions centered around $r_e=5.1$ \AA\ is
completely inadequate for representing such dramatically expanded
geometries. 


\begin{figure}
\begin{center}
\epsfig{file=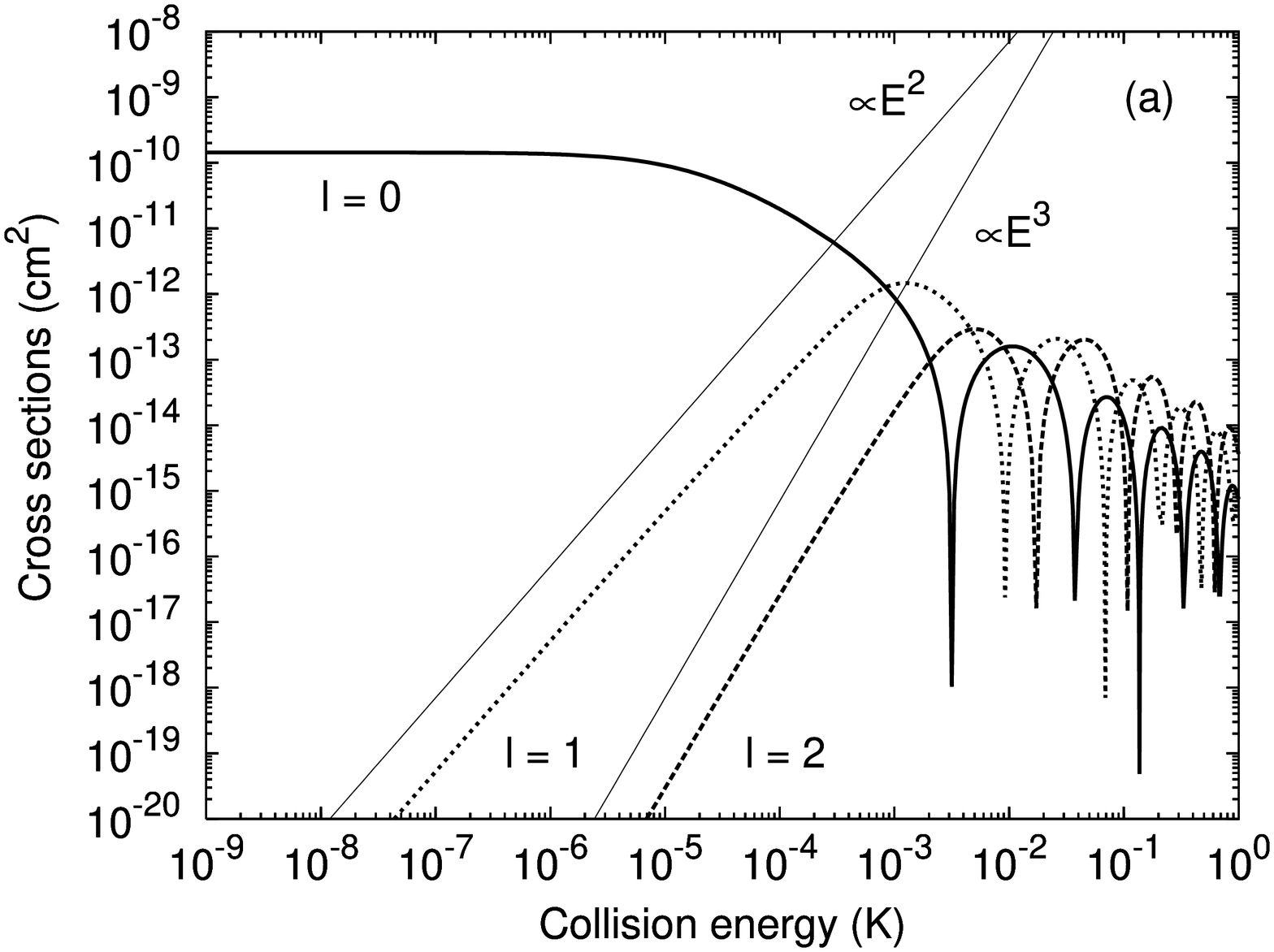,angle=-90,width=85mm}
\epsfig{file=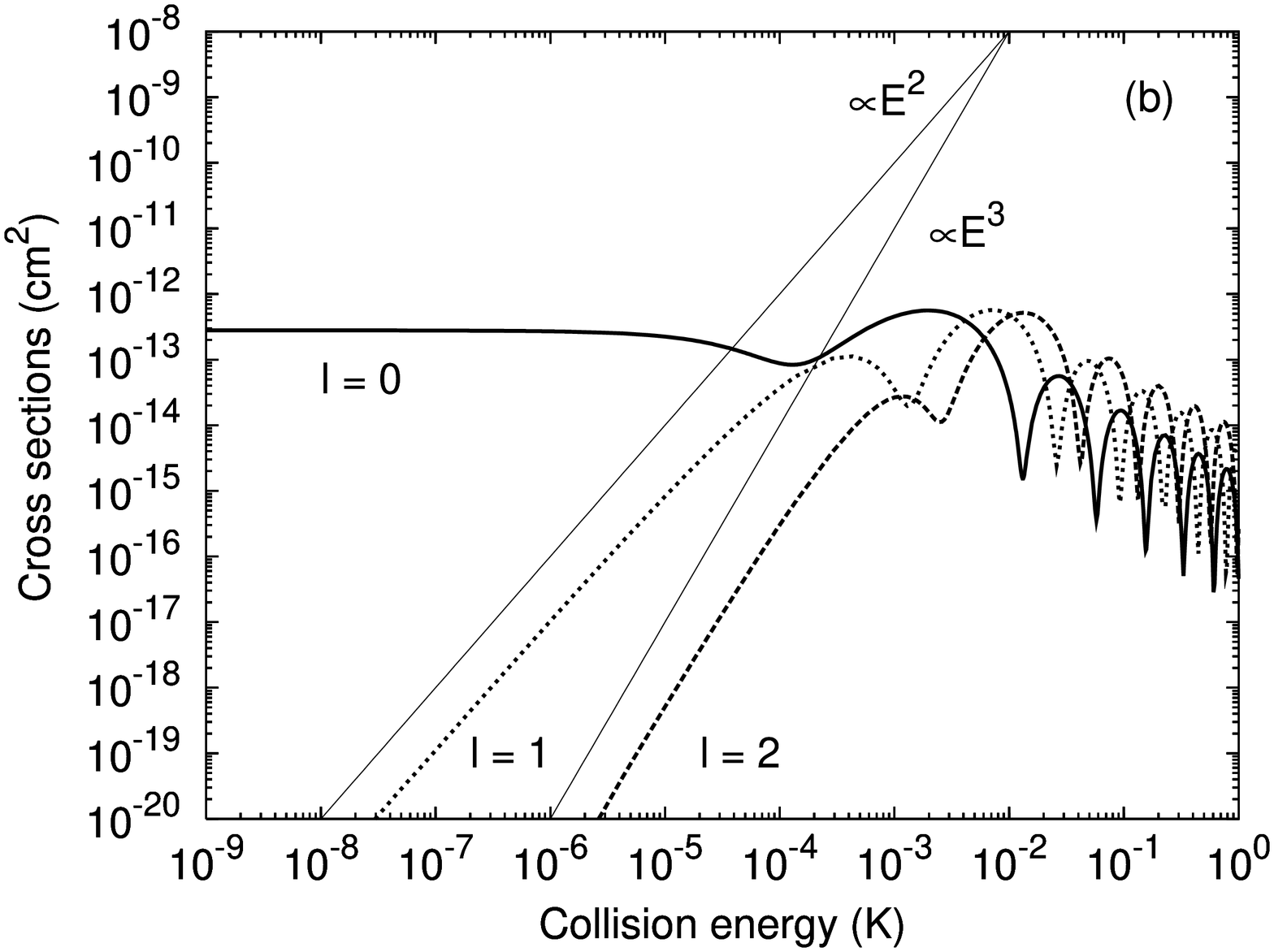,angle=-90,width=85mm}
\end{center}
\caption{
Contributions to elastic cross sections from s, p and d waves for 
collinear Na + Na$_2$ ($v=1$) collisions,
neglecting reactive channels.
(a) additive potential.
(b) non-additive potential.
}
\label{fig1}
\end{figure}

\begin{figure}
\begin{center}
\epsfig{file=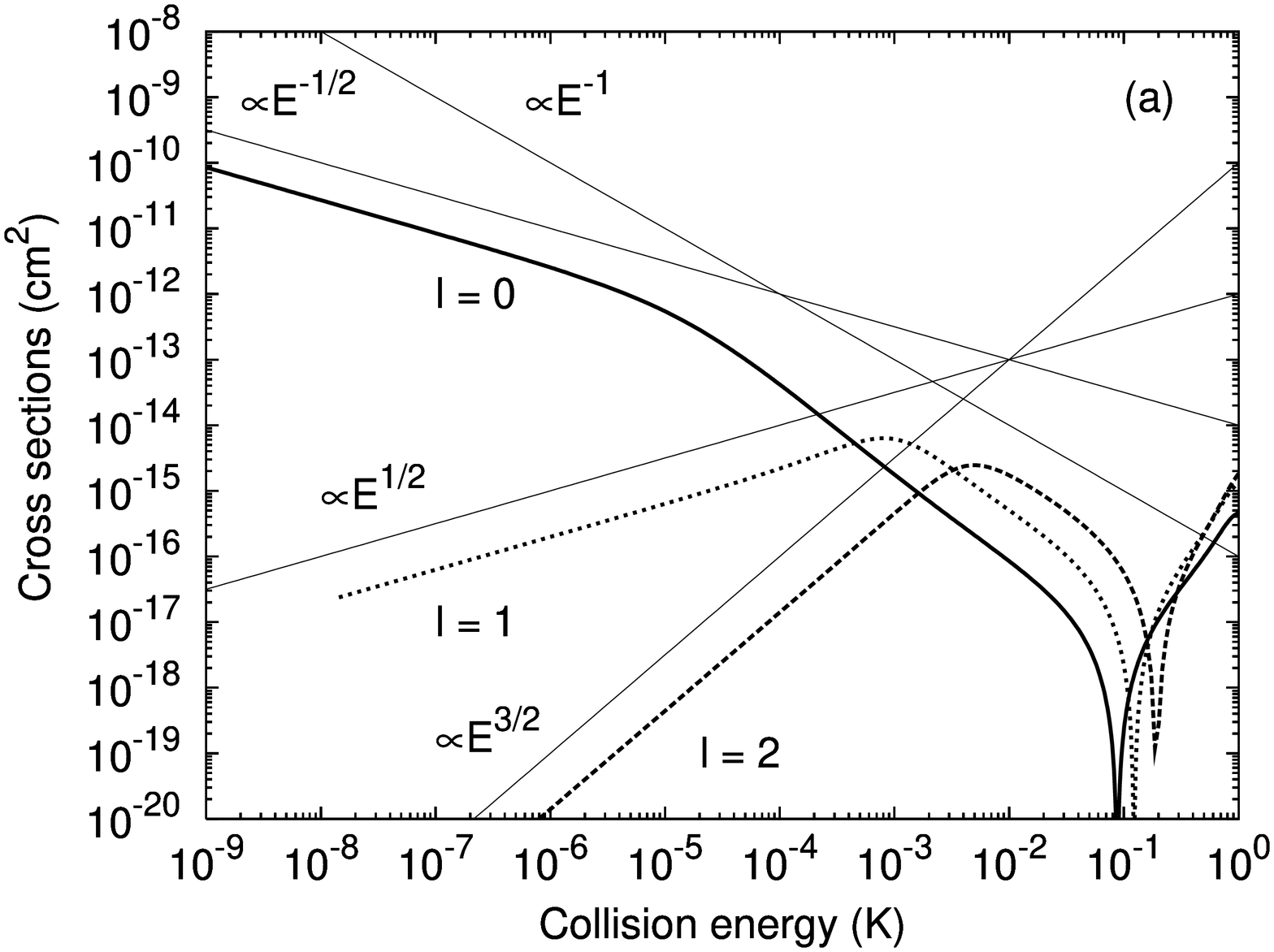,angle=-90,width=85mm}
\epsfig{file=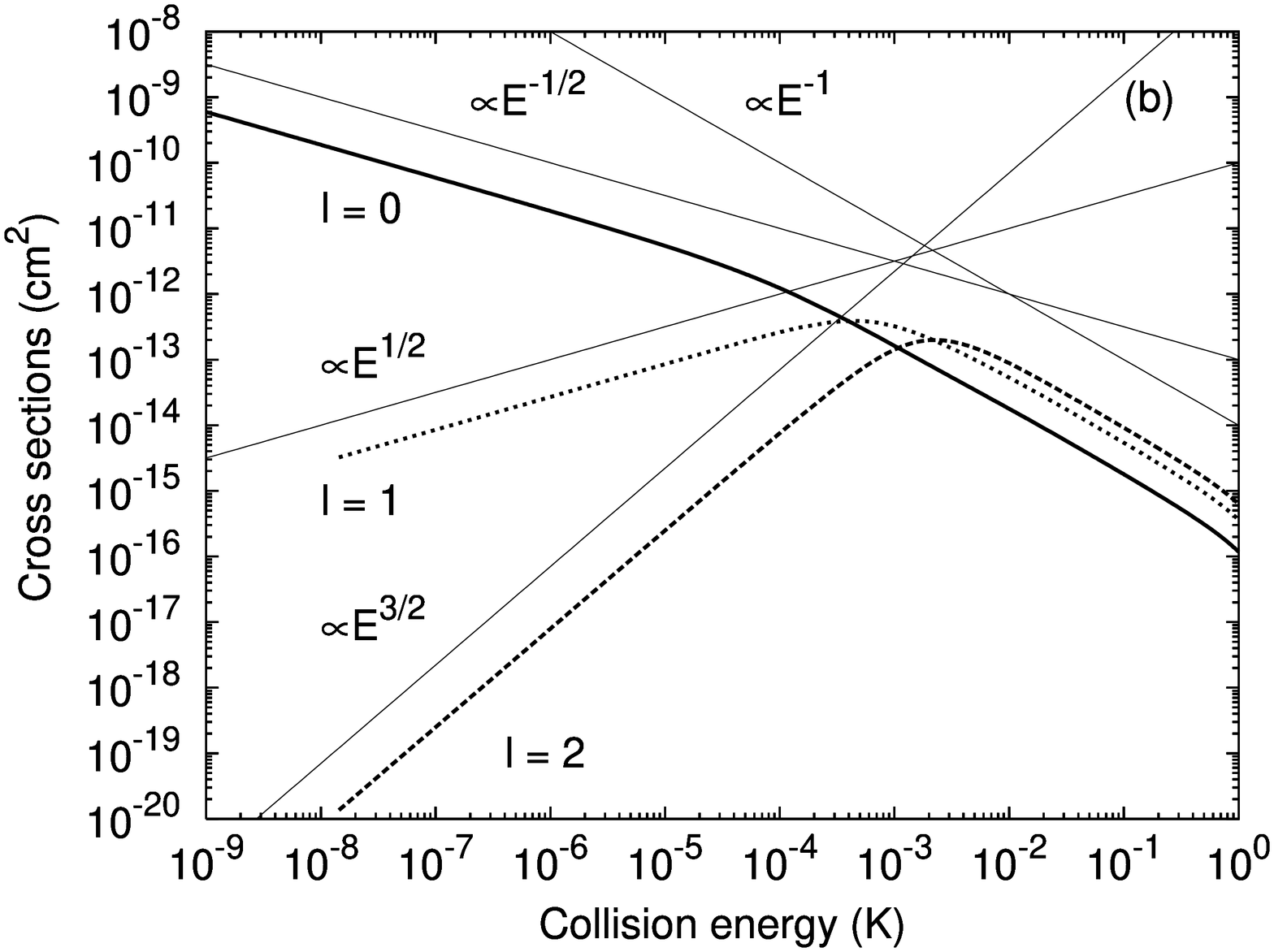,angle=-90,width=85mm}
\end{center}
\caption{
Contributions to inelastic cross sections (to produce Na$_2$ ($v=1$) 
from s, p and d waves for collinear Na + Na$_2$ ($v=1$) collisions, 
neglecting reactive channels.
(a) additive potential.
(b) non-additive potential.
}
\label{fig2}
\end{figure}

This problem does not arise for collinear scattering, for which the
vibrational basis set is adequately converged for $v_{\rm max}=20$. We
therefore carried out collinear calculations of the elastic and vibrational
relaxation cross sections. The coupled equations were integrated from
$R=2.3$ \AA\ to 2500 \AA. The resulting elastic and inelastic cross
sections for collisions of Na$_2$ initially in $v=1$ are shown in Figs.\
\ref{fig1} and \ref{fig2} for both the pairwise-additive and non-additive
potentials.

The cross sections all follow the expected Wigner threshold laws
\cite{Bal97thresh,Wig48,Rjpb00} at energies below $10^{-6}$~K. For elastic
collisions, the cross sections are proportional to $E^0$, $E^2$ and $E^3$
for s, p and d waves respectively. The $E^3$ dependence for the d wave
arises because the long-range dispersion ($R^{-6}$) term in the atom-diatom
potential modifies the threshold behavior for $l>1$ \cite{Rjpb00}. For
vibrational relaxation, the cross sections are proportional to $E^{-1/2}$,
$E^{1/2}$ and $E^{3/2}$ for s, p and d waves respectively. The oscillations
that occur for $E>1$ mK arise simply from zeroes in $\sin^2\eta$, where
$\eta$ is the scattering phase shift.

The {\it elastic} cross sections are dramatically different for the
additive and non-additive potentials (by a factor of almost $10^3$), while
the {\it inelastic} cross sections are much more similar (differing by a
factor of only 10). The large difference in the elastic cross sections
presumably occurs because the zero-energy elastic cross section may be
written as $\sigma^{\rm el}_0 = 4\pi |a|^2$, where $a$ is the atom-diatom
complex scattering length \cite{Bal97complexa}. This is accidentally close
to zero for the non-additive potential: $a = (-0.90-1.19i)$ nm, compared to
$(33.83 -0.46i)$ nm for the additive potential. This has the effect that
$\sigma^{\rm el} > \sigma^{\rm inel}$ for $T>10^{-10}$~K on the additive
surface, but only for $T>10^{-4}$~K on the non-additive surface.

An intriguing feature of the results is the appearance of a second linear
region (in the log-log plots) for inelastic s-wave scattering between about
$10^{-5}$ and 10$^{-2}$~K. The cross section in this region is proportional
to $E^{-1}$ for the non-additive potential but to about $E^{-4/3}$ for the
additive potential. A possible reason for this is that the vibrational
coupling at long range decays as $R^{-6}$ for the full potential but as
$R^{-8}$ for the pairwise-additive potential.

As described above, the MOLSCAT calculations are restricted to collinear
geometries and a single arrangement channel. To go beyond this requires a
reactive scattering formalism in which all three arrangement channels are
included. Such calculations are more expensive, and have not previously
been attempted for systems containing three non-hydrogen atoms.

As a first step towards a rigorous treatment, we have performed
three-dimensional quantum reactive scattering calculations for total
nuclear orbital angular momentum $J=0$. The configuration space is divided
into an inner and an outer region, depending on the atom-diatom distance.
In the inner region, we use a formalism based on body-frame democratic
hyperspherical coordinates \cite{launay89}. This has already proved
successful in describing atom-diatom insertion reactions such as N($^2$D) +
H$_2$$\rightarrow$ NH + H \cite{nh2jcp99} and O($^1$D) + H$_2$$\rightarrow$
OH + H \cite{oh2jcp01,oh2prl01}. A related approach has also been used in
studies of three-body recombination of ultracold atoms \cite{Esr99recomb}.
The scattering wave function is expanded on a set of hyperspherical
adiabatic states. This yields a set of close-coupling equations, which in
our method are solved using the Johnson-Manolopoulos log-derivative
propagator \cite{mano86}. In the outer region, we use the Arthurs-Dalgarno
formalism \cite{arthurs60}, which is based on Jacobi coordinates. Matching
of the wavefunctions in the inner and outer regions is performed on a
boundary which is an hypersphere of radius 25~\AA. This yields the
reactance $K$-matrix and the scattering $S$-matrix.

The inner region starts at a hyperradius of 4 \AA\ and is split into 297
sectors. The adiabatic states in each sector are obtained by a variational
expansion on a basis of hyperspherical harmonics with $A_1$ symmetry. They
are fully symmetric with respect to particle permutations to account for
the indistinguishability of atoms. At large hyperradius, the adiabatic
states concentrate into the arrangement channels and describe Na$_2$
molecules in even $j$ states. At small hyperradius, they span a large
fraction of configuration space and allow for atom exchange. The
hyperspherical harmonic basis is truncated at $\Lambda_{\rm max}$, the
maximum value of the grand angular momentum. $\Lambda_{\rm max}$ varies
from 198 (867 harmonics) at small hyperradius to 398 (3400 harmonics) at
large hyperradius. A fixed number of 135 adiabatic states is used in the
close-coupling expansion in each sector. At the boundary between the inner
and outer regions, the adiabatic states are projected onto a set of Na$_2$
rovibrational states, with $j_{\rm max} = 48$, 44, 40, 36, 30, 26, 20, 10
for vibrational levels $v=0,1,\dots,7$. The boundary between the inner and
outer regions was placed at a distance such that couplings due to the
atom-diatom residual interaction can be neglected outside the boundary. In
the outer region, regular and irregular solutions of a radial
Schr{\"o}dinger equation which includes the isotropic ($R^{-6}$) part of
the interaction were integrated inwards from very large distances (5000
\AA).

\begin{figure}
\begin{center}
\epsfig{file=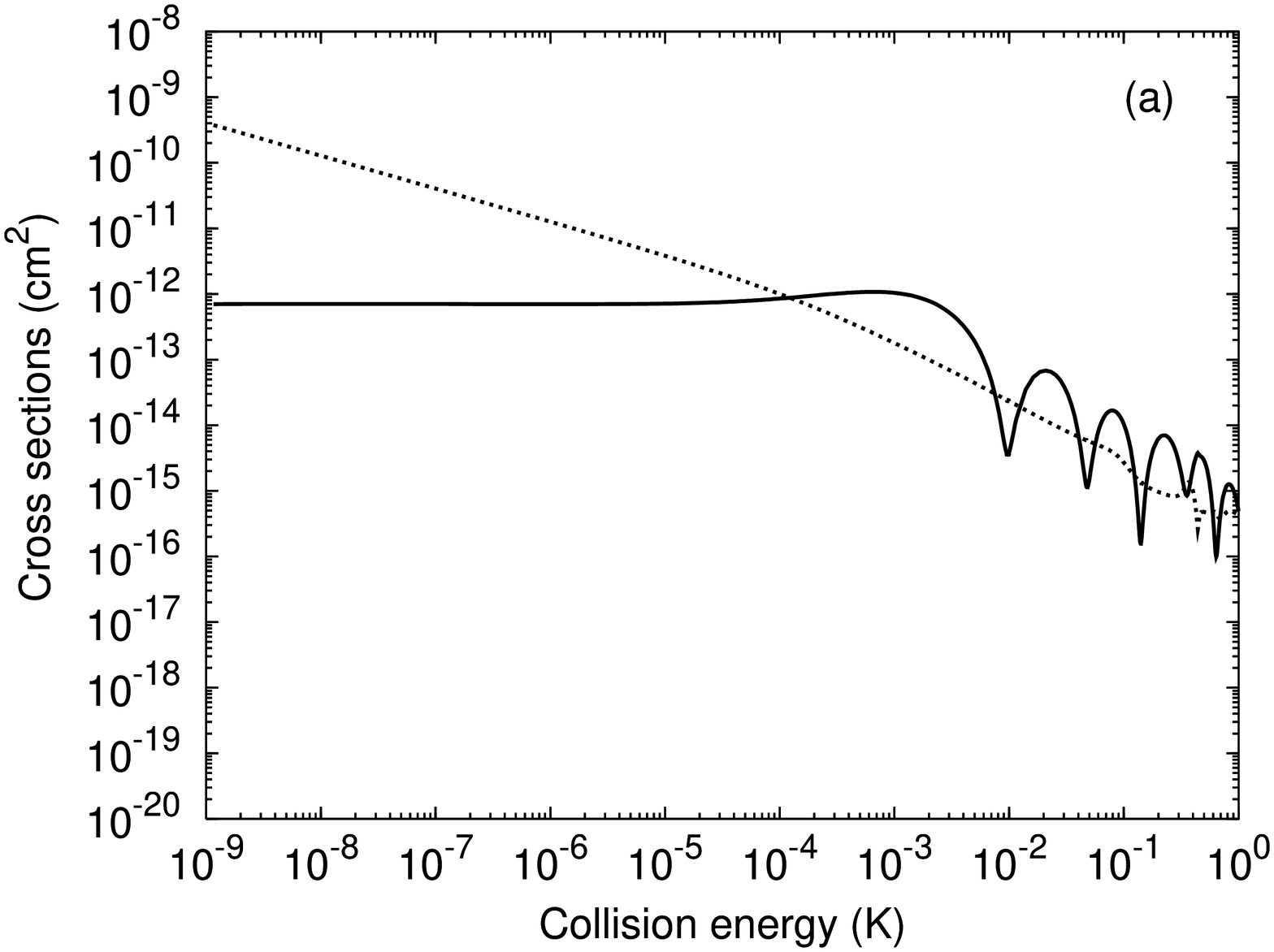,angle=-90,width=85mm}
\epsfig{file=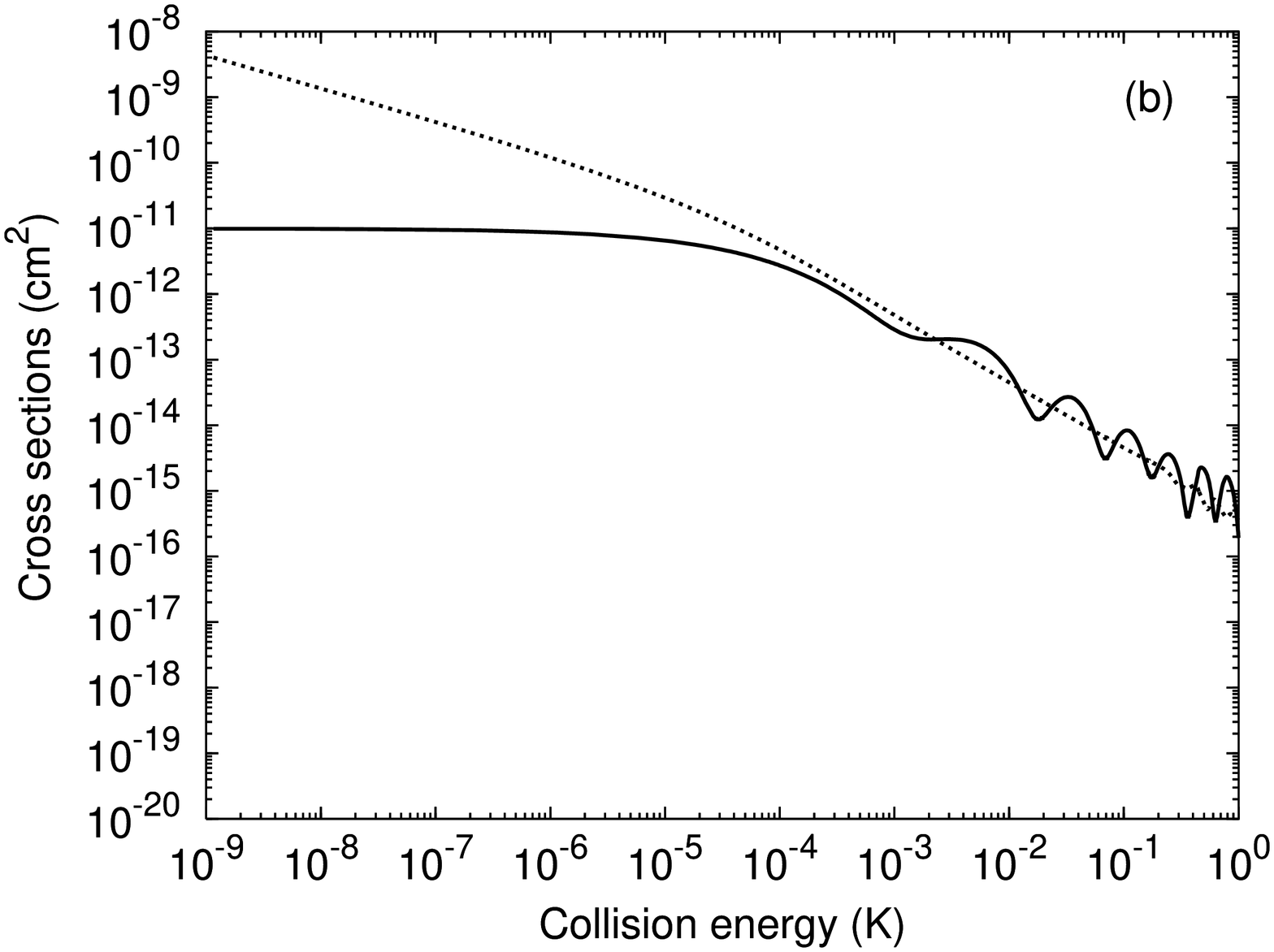,angle=-90,width=85mm}
\end{center}
\caption{
Cross sections from quantum reactive scattering calculations 
for Na + Na$_2$($v=1,j=0$) (s-wave scattering).
Elastic and quenching results are shown as
solid and dotted lines.
(a) additive potential.
(b) non-additive potential.
}
\label{fig3}
\end{figure}

Fig.\ \ref{fig3} shows the cross sections as a function of collision energy
for the additive and non-additive potentials. All collisions that are
energetically elastic, with or without atom exchange, are included in the
elastic cross section. All other processes (which produce Na$_2$ $(v=0,j)$)
contribute to the quenching cross section. Na$_2$ rotational levels up to
$j=20$ are energetically accessible at the energy of the $v=1$ state (23.5
cm$^{-1}$), and all accessible levels are populated in the products.

It may be seen that the Wigner threshold laws for the elastic and quenching
cross sections hold below $10^{-5}$~K for both potentials. The cross
sections are larger for the non-additive than for the additive potential,
by a factor of about 10 for both elastic and quenching cross sections. The
$E^{-1/2}$ dependence of the quenching cross sections corresponds to a
constant rate coefficient below $10^{-5}$~K, which is $k^{\rm inel} = 4.8
\times 10^{-11}$ cm$^3$ s$^{-1}$ for the additive potential and $5.2 \times
10^{-10}$ cm$^3$ s$^{-1}$ for the non-additive potential. The corresponding
scattering lengths are $a = (2.26-0.80i)$ and $(2.38-8.54i)$ nm,
respectively.

Outside the Wigner region, the cross sections have a more complicated
energy dependence. The quenching probability increases with increasing
energy and approaches unity at the limit of the Wigner region. The
quenching cross sections thus vary approximately as $E^{-1}$ above
$10^{-4}$~K, because of the $k^{-2}$ factor in the expression for
the cross section. Above $10^{-3}$~K, the elastic cross sections show
oscillations which are similar to but less pronounced than the ones in
Fig.\ \ref{fig1}.

Finally, the ratio of quenching to elastic cross sections is larger than 1
at energies below $10^{-4}$~K for the additive potential and below
$10^{-3}$~K for the non-additive potential. It increases up to 500 in the
nK range for both potentials.

In future work, we intend to investigate the dependence of quenching
rates on the initial vibrational quantum number and to investigate the
effects of magnetic fields and nuclear spin coupling.

The three-dimensional quantum dynamical calculations were performed on a
NEC-SX5 vector supercomputer, through a grant from the ``Institut du
D{\'e}veloppement des Ressources en Informatique Scientifique'' (IDRIS) in
Orsay (France). PS is grateful to the EPSRC for a Research Associateship 
under grant no.\ GR/R17522/01.

\end{document}